\documentclass[reprint,prb,superscriptaddress,amsmath,amssymb,aps,twocolumn]{revtex4-1}
\usepackage{color}
\usepackage{amsmath}
\usepackage{amsfonts}
\usepackage{amssymb}
\usepackage{MnSymbol}
\usepackage{graphicx}
\usepackage[caption=false]{subfig}
\usepackage{tikz}
\usetikzlibrary{arrows}

\usepackage{tikz-feynman}
\tikzfeynmanset{compat=1.0.0}

\begin{document}
\title{Control of open quantum systems: Manipulation of a qubit
coupled to a thermal bath by an external driving field.}
\author{Haoran Sun}
\affiliation{Department of Chemistry \& Biochemistry, University of California San Diego, La Jolla, CA 92093, USA}
\author{Michael Galperin}
\email{migalperin@ucsd.edu}
\affiliation{Department of Chemistry \& Biochemistry, University of California San Diego, La Jolla, CA 92093, USA} 

\begin{abstract}
Fast and reliable manipulation with qubits is fundamental for any quantum technology. The implementation of these manipulations in physical systems is the focus of studies involving optimal control theory. Realistic physical devices are open quantum systems. So far, studies in optimal control theory have primarily utilized the Redfield/Lindblad quantum master equation to simulate the dynamics of such systems. However, this Markov description is not always sufficient. Here, we present a study of qubit control utilizing the nonequilibrium Green's function method. We compare the traditional master equation with more general Green's function results and demonstrate that even in the parameter regime suitable for the application of the Redfield/Lindblad approach, the two methods yield drastically different results when addressing evolution involving mixed states. In particular, we find that, in addition to predicting different optimal driving profiles, a more accurate description of system evolution enables the system to reach the desired final state much more quickly. We argue that the primary reason for this is the significance of the non-Markov description of driven system dynamics due to the effect of time-dependent driving on dissipation.
\end{abstract}

\maketitle

%%%%%%%%%%%%%%%%%%%%%%%%%%%%%%%%%%%%%%%%%%%%%%%%%%%%%%%%%%%%%%%%%%%%%%%%%%%%%%%
\section{Introduction}\label{intro}
The capability to create, transfer, and read quantum information forms the basis of 
quantum measurement~\cite{guhne_colloquium_2023}, 
quantum metrology~\cite{toth_quantum_2014},
and quantum computing~\cite{adesso_measures_2016,hu_quantum_2018}. 
Corresponding methods and limits of information manipulation are explored 
in research on quantum steering~\cite{uola_quantum_2020}, 
quantum resource theories~\cite{chitambar_quantum_2019}, 
and quantum thermodynamics~\cite{goold_role_2016,vinjanampathy_quantum_2016,kosloff_quantum_2019}. 
Notably, the rapid and reliable reset of a qubit is a crucial prerequisite for any 
quantum technology~\cite{palao_quantum_2002,kallush_scaling_2011}. 
The implementation of these manipulations in physical systems is the focus of 
studies in quantum optimal control theory (OCT)~\cite{glaser_training_2015,deffner_quantum_2017,bukov_reinforcement_2018,frisk_kockum_ultrastrong_2019,guery-odelin_shortcuts_2019,koch_quantum_2022}.

Optimal control addresses the challenge of achieving a desired quantum state of 
a system in the most efficient manner. Typically, this control is implemented using 
time-dependent external fields. Initially, optimal control theory concentrated 
on isolated quantum systems, where the Schr{\" o}dinger equation governed 
the system's evolution~\cite{peirce_optimal_1988,kosloff_wavepacket_1989,szakacs_locking_1994,palao_optimal_2003,koch_stabilization_2004,palao_protecting_2008,tomza_optimized_2012,aroch_optimizing_2018,schaefer_optimization_2020}. 
Effective computational algorithms for attaining the desired outcomes in such 
systems are documented in the literature~\cite{ndong_chebychev_2009,rossignolo_quocs_2023}.

Realistic experimental setups involve open quantum systems. 
The need to consider the open nature of the system has led to modifications in 
OCT studies. Initially, a surrogate Hamiltonian approach was proposed~\cite{baer_quantum_1997,gelman_minimizing_2005,asplund_optimal_2011,asplund_optimal_2012,abdelhafez_gradient-based_2019}. 
In this approach, the Schr{\" o}dinger equation is applied to a finite surrogate 
Hamiltonian that substitutes the physical (infinite) system-bath description. 
This finite representation is crafted to produce reasonable short-time dynamics 
of the system coupled to an infinite bath.

To simulate longer trajectories and incorporate accurate thermodynamic behavior, 
the open character of the system should be considered more precisely. 
This led to the substitution of Schr{\" o}dinger with the quantum master equation (QME) 
as the dynamical law in OCT studies. Since optimization relies on an iterative 
procedure, QME is typically employed in its most basic (Redfield/Lindblad) form. Following tradition, OCT considerations with QME as the dynamical law 
overlook the influence of the time-dependent field on the dissipator~\cite{bartana_laser_1993,tang_generalized_1996,tannor_laser_1999,ohtsuki_monotonically_1999,bartana_laser_2001,xu_optimal_2004,katz_decoherence_2007,basilewitsch_beating_2017,basilewitsch_quantum_2019,aroch_employing_2023}. 
Subsequently, to enhance modeling realism, the effect of the field on 
the dissipator was incorporated within the instantaneous (adiabatic) 
approximation~\cite{dann_quantum_2021,kallush_controlling_2022}.

Markov evolution simulated within the Redfield/Lindblad QME
(similar to other weak coupling theories~\cite{lostaglio_introductory_2019})
is not always accurate in predicting open system responses~\cite{esposito_efficiency_2015,gao_simulation_2016}.
Thus, non-Markov OCT was performed for the spin-boson system
using the hierarchical equation of motion approach (HEOM)~\cite{mangaud_non-markovianity_2018}.
HEOM is a highly promising technique for simulating time-dependent
and transient processes due to its linear scaling with time.
Early versions of HEOM were restricted to high temperatures
and specific band structures. Subsequent developments allow
for trading some of its advantages for significantly improved access to
more general band structures and lower temperatures~\cite{erpenbeck_extending_2018}.
Nevertheless, it remains unclear whether experimentally relevant 
low temperatures are achievable with this technique.

Green's function methods~\cite{cohen_greens_2020} 
are capable of overcoming the limitations of
the techniques used in OCT literature so far.
Here, we employ the nonequilibrium Green's function (NEGF) method
in OCT studies of a qubit. 
Following Ref.~\onlinecite{kallush_controlling_2022}
we consider optimization of laser pulse with goals of 
heating, cooling, and particular state preparation
of an open quantum system. 
We compared QME results with more general NEGF simulations.
We note that  there are two aspects to quantum control analysis:
1.~the optimal control method and 2.~the theoretical approach used to describe 
the evolution of the system. In our study we follow the optimal control method
employed in Ref.~\onlinecite{kallush_controlling_2022}.
While more advanced optimization methods are available in OCT literature,
focus of our consideration is description of the evolution of the system.
We suggest the NEGF approach exemplified here and~\cite{esposito_efficiency_2015,gao_simulation_2016}
in non-Markov scenarios of strong driving/coupling, where
the usual Markov (Redfield/Lindblad-type) approaches become inaccurate. In
those cases, NEGF could be a way around solving the
non-Markov integro-differential equations with memory kernels. 
Besides preserving complete positivity of the system density operator~\cite{hyrkas_cutting_2022},
NEGF is capable of proper description of system-bath interaction including
quantum effects (bath induced coherences) and influence of time-dependent
driving on the coupling.
 In this sense, choice of optimization
procedure is of secondary importance.

The structure of the paper is as follows. 
Section~\ref{model} introduces the model and describes the methods 
used in the simulations. A comparison between the numerical results of 
QME and NEGF simulations is presented in Section~\ref{numres}. 
Section~\ref{conclude} summarizes our findings.

%%%%%%%%%%%%%%%%%%%%%%%%%%%%%%%%%%%%%%%%%
\begin{figure}[htbp]
    {\centering\includegraphics[width=0.8\linewidth]{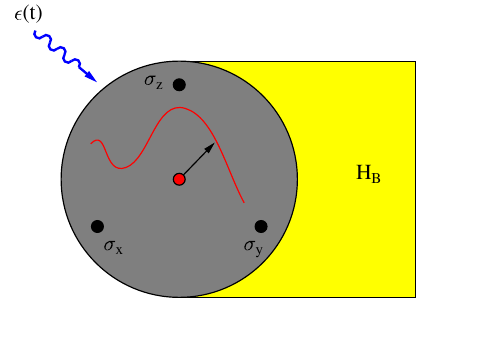}}
    \caption{\label{fig1}
    (Color online)
    Sketch of a qubit coupled to a thermal bath and driven by
    external field.}
\end{figure} 
%%%%%%%%%%%%%%%%%%%%%%%%%%%%%%%%%%%%%%%%%
%%%%%%%%%%%%%%%%%%%%%%%%%%%%%%%%%%%%%%%%%

\section{Control of qubit coupled to bath} \label{model}
Here, we introduce the model, describe the methods employed in the simulations 
of the system dynamics, and discuss the optimization procedure.

\subsection{Model}
Following Ref.~\onlinecite{kallush_controlling_2022}
we consider a qubit $S$ coupled to thermal bath $B$
and driven by external time-dependent field $\epsilon(t)$
(see Fig.~\ref{fig1})
\begin{equation}
\label{H}
    \hat H(t) = \hat H_S(t) + \hat H_B + \hat H_{SB}
\end{equation}
where $\hat H_S(t)$ and $\hat H_B$ describe decoupled system and bath,
respectively.
$\hat H_{SB}$ introduces interaction between them.
Explicit expressions are
(here and below $\hbar=k_B=1$)
\begin{align}
\label{HS}
%    \hat H_S(t) &= \Delta\left(1+\hat s_x\right) 
%     +\epsilon(t)\,\hat s_z
\hat H_S(t) &= \sum_{i=1}^{2} \Delta\,\hat n_i
+\frac{\Delta}{2}\left(\hat d^\dagger_1\hat d_2+\mbox{H.c.}\right)
\\ &
+\frac{\epsilon(t)}{2}\left(\hat n_2-\hat n_1\right)
 \nonumber    \\
\label{HB}
    \hat H_B &= \sum_\alpha \omega_\alpha\left(\hat a_\alpha^\dagger \hat a_\alpha + \frac{1}{2}\right)
    \\
\label{HSB}
%    \hat H_{SB} &= \hat s_y \sum_\alpha g_\alpha(\hat a_\alpha + \hat a_\alpha^\dagger)
\hat H_{SB} &= \frac{i}{2}\left(\hat d_2^\dagger\hat d_1-\hat d_1^\dagger\hat d_2\right)
\sum_\alpha g_\alpha\left(\hat a_\alpha+\hat a_\alpha^\dagger\right)
\end{align}
Here, $\hat d_i^\dagger$ ($\hat d_i$) and $\hat a_\alpha^\dagger$ ($\hat a_\alpha$) 
creates (annihilates) electron in level $i$ ($i=1,2$) and excitation in
mode $\alpha$ of thermal bath, respectively. 
$\hat n_i\equiv \hat d_i^\dagger\hat d_i$ is the particle number operator.
Driving field $\epsilon(t)$ is modeled
as set of harmonic modes acting within a finite time interval
\begin{equation}
\label{epsilon}
    \epsilon(t) = 
    \exp\left[-\left(\frac{t-t_c/2}{t_c}\right)^2\right] 
    \sum_{k=1}^{M} c_k\sin\left( \nu_k t\right)
\end{equation}   
where, $M$ is the number of harmonic modes in the signal and
$t_c$ is the control time.

Note that the Hamiltonian commutes with the system total particle number operator
\begin{equation}
\left[\hat H(t),\hat n_1+\hat n_2\right]=0
\end{equation}
Thus, the formulation properly describes qubit (spin-1/2)
evolution when confined to the one-particle subspace.

%%%%%%%%%%%%%%%%%%%%%%%%%%%%%%%%%%%%%%%%%

\subsection{QME formulation}
Following Ref.~\onlinecite{kallush_controlling_2022}, for the standard OCT 
treatment, we simulate the system's evolution using the Redfield/Lindblad QME. 
We utilize an advanced version of the QME that accounts for the effect of 
a time-dependent driving field on the dissipation superoperator~\cite{dann_time-dependent_2018}. 
The corresponding Markov EOM for the system's density operator in the interaction picture is (see Appendix~\ref{appA} for details)
\begin{align}
\label{QME}
    \frac{d}{dt}\hat{\tilde\rho}_S(t) &= \sum_m\bigg\{
    -i\,\mbox{Im}\left(\Gamma_m(t)\right)
    \left[\hat F_m^\dagger(t)\hat F_m(t), \hat{\tilde\rho}_S(t)\right]
    \nonumber
    \\ & \qquad
    + 2\,\mbox{Re}\left(\Gamma_m(t)\right)
    \bigg[ \hat F_m(t)\hat{\tilde\rho}_S(t)\hat F_m^\dagger(t) 
    \\ &\qquad\qquad\qquad\qquad
    - \frac{1}{2}\{\hat F_m^\dagger(t)\hat F_m(t),
   \hat{\tilde\rho}_S(t)\}
    \bigg]\bigg\}
     \nonumber 
\end{align}
Here, $\hat F_m(t)$ and $\Gamma_m(t)$ are the jump operators and 
time-dependent coefficients defined in Eqs.~(\ref{appA_defF}) 
and (\ref{appA_defGamma}), respectively.

%%%%%%%%%%%%%%%%%%%%%%%%%%%%%%%%%%%%%%%%%

\subsection{NEGF formulation}
Single particle Green's function
\begin{equation}
\label{GF}
G_{ij}(\tau_1, \tau_2) = -i\langle{T_c\, d_{i}(\tau_1)\, d_{j}^\dagger(\tau_2)}\rangle
\end{equation}
is the central object to describe dynamics of the system within the NEGF. 
Here, $\tau_{1,2}$ are the Keldysh contour variables,
$T_c$ is the contour ordering operator, and
$\hat d_i^\dagger(\tau)$ ($\hat d_i(\tau)$) is the electron creation (annihilation)
operator in the Heisenberg picture.
Single particle density matrix is given by the lesser projection
of the Green's function 
\begin{equation}
\label{rho}
    \rho_{ij}(t) = -i\, G_{ij}^<(t, t)
\end{equation}
Within the NEGF, dynamics of the system is governed by 
the Kadanoff-Baym equation
\begin{align}
\label{KB}
 i\frac{\partial}{\partial\tau_1} G_{ij}(\tau_1, \tau_2) &= 
 \delta_{i,j}\,\delta(\tau_1,\tau_2) 
     \nonumber
 \\ &
 + \sum_m \left[H_S(t_1)\right]_{im}\, G_{mj}(\tau_1,\tau_2)
 \\ &
 + \sum_m\int_c d\tau\,\Sigma_{im}(\tau_1,\tau)\, G_{mj}(\tau,\tau_2)
      \nonumber
\end{align}
Here, $t_1$ is physical time corresponding to contour variable $\tau_1$.
$\Sigma$ is the self-energy which describes the influence of the bath.
Its explicit expression within the Hartree-Fock approximation is
given in Eqs.~(\ref{appB_SE})-(\ref{appB_SEF}).

%%%%%%%%%%%%%%%%%%%%%%%%%%%%%%%%%%%%%%%%%

\subsection{Control tasks}
Dynamical trajectories generated within QME or NEGF are used
in an iterative procedure to reach control tasks.
Below we consider three such tasks: reaching a particular quantum state (reset),
heating, and cooling of the system. 

The tasks are achieved
by optimization of the driving force profile $\epsilon(t)$.
Specifically, we use coefficients $c_k$ 
in the Fourier series (\ref{epsilon}) as adjustable parameters,
while frequencies $\nu_k$ are fixed
\begin{equation}
    \nu_k = \frac{k\pi}{t_c},~ k = 1,2,3,\dots
\end{equation}

We start with a decoupled qubit and bath with the qubit being in
an initial state $\hat\rho_i=\hat\rho(t=0)$. 
Within a predefined finite control time $t_c$, we search for such driving
force profile $\epsilon(t)$ that by the time $t=t_c$ the qubit evolves to some
desired state $\hat\rho_f\approx\hat\rho(t_c)$.
Note that the density operator of a qubit can be decomposed as
\begin{equation}
 \hat\rho = \left(\frac{1}{2} + x\, \hat s_x + y\,\hat s_y + z\,\hat s_z\right)
\end{equation}
Here, $\hat s_{x,y,z}$ are the spin-1/2 operators.
That is, state of the qubit is fully defined by vector $(x,y,z)$ and can be represented
as a point within 3D unit sphere, while distance between two states
can be defined as distance between a pair of 3D vectors.

Quality of the control is defined by a cost function 
$R[\hat\rho_i,\epsilon(t)]$. 
That is, goal of the optimization is
finding functions $c_k(t)$ which minimize the cost function.
In particular, for reset (r) and heating (h) tasks,
the cost function is defined as the distance 
between the state of the system at the end of the trajectory, 
$\hat\rho(t_c)$, and desired outcome $\hat\rho_f$
\begin{equation}
R_{r,h}[\hat\rho_i,\epsilon(t)] \equiv \|\hat\rho_f-\hat\rho(t_c)\|
\end{equation} 
The cost function for the cooling (c) task is defined as
\begin{equation}
R_{c}[\hat\rho_i,\epsilon(t)] \equiv -\|\hat\rho(t_c)\|{}^2
\end{equation}
This form is used because of its faster convergence.

%%%%%%%%%%%%%%%%%%%%%%%%%%%%%%%%%%%%%%%%%
%%%%%%%%%%%%%%%%%%%%%%%%%%%%%%%%%%%%%%%%%

\section{Numerical results and discussion}\label{numres}
Time evolution within the QME formulation is numerically simulated using 
the Runge-Kutta method of fourth order (RK4)~\cite{press_numerical_1997}.
Time propagation within the NEGF formulations is simulated 
following a scheme introduced in Ref.~\onlinecite{stan_time_2009}.
The NEGF simulations employ Fast Fourier transform (FFT)
which is numerically performed utilizing the FFTW library~\cite{FFTW}.
Differential evolution~\cite{storn_differential_1997}
is employed as an optimization algorithm
to find driving force profile for both QME
and NEGF dynamics, the same random generator,
iterative depth and population size
are used in both cases.

Below we compare results of simulations within QME and NEGF
for the optimized driving force $\epsilon(t)$, entropy production $S_i(t)$,
entropy $S(t)$, and cost function $R[\hat\rho_i,\epsilon(t)]$.
Entropy is obtained by integrating differential form of  the second law of 
thermodynamics
\begin{equation}
\label{2law}
 \frac{d}{dt} S(t) = \beta {\dot Q}_B(t) + {\dot S}_i(t)\qquad
  {\dot S}_i(t)\geq 0
\end{equation}
where ${\dot Q}_B(t)$ and $ {\dot S}_i(t)$ are heat flux between
the qubit and thermal bath and entropy production rate, respectively.
Heat flux is defined as rate of energy change in the bath
\begin{equation}
 {\dot Q}_B(t) = -\sum_\alpha\omega_\alpha\,\frac{d}{dt} \langle\hat a_\alpha^\dagger(t)\hat a_\alpha(t)\rangle
\end{equation}

Expressions for entropy and entropy production depend on formulation of the 
second law. 
For systems weakly coupled to their baths (QME formulation)
the former is given by the von Neumann expression 
\begin{equation}
S(t) = -\mbox{Tr}_S\left\{\hat\rho_S(t)\,\ln\hat\rho_S(t)\right\}
\end{equation}
which together with 
expression for heat flux yields entropy production rate~\cite{ptaszynski_thermodynamics_2019}
\begin{equation}
\begin{split}
{\dot Q}_B(t) &= \mbox{Tr}_S\left\{\left(\mathcal{L}_D\rho_S(t)\right)\,\hat H^S(t)\right\}
\\
{\dot S}_i(t) &= -\mbox{Tr}_S\left\{\left(\mathcal{L}_D\rho_S(t)\right)\,\ln\hat\rho_S(t)\right\}
\end{split}
\end{equation}
Here, $\mathcal{L}_D$ is dissipator part of the Liouvillian defined by
the  last two lines of Eq.(\ref{QME}). 

For systems strongly coupled to their baths we follow formulation
of Ref.~\onlinecite{zhou_quantum_2024}. Expressions for heat flux
and entropy production rate are
\begin{align}
 & {\dot Q}_B(t) = \int_0^\infty \frac{d\omega}{2\pi}\, \omega\, i_B(t,\omega)
 \\
& {\dot S}_i(t) = \int_0^\infty \frac{d\omega}{2\pi}\,\left\{
 \phi^{out}(t,\omega)\left[\ln\phi^{out}(t,\omega)-\ln\phi^{in}(\omega)\right]
\right. \nonumber \\ & \left.
 -\left(1+\phi^{out}(t,\omega)\right)\left[\ln\left(1+\phi^{out}(t,\omega)\right)-
 \ln\left(1+\phi^{in}(\omega)\right)\right]
 \right\}
 \nonumber
\end{align}
Here, $i_B(t,\omega)$ is the energy resolved particle (phonon)
flux between the system and thermal bath defined in Eq.~(\ref{appC_iB}),
$\phi^{in}(\omega)$ and $\phi^{out}(t,\omega)$ are the thermal population
of incoming and non-thermal population of outgoing states 
in the bath. Explicit expressions for the current and populations
in terms of the Green's function (\ref{GF}) are given in Appendix~\ref{appC}. 

The parameters of the simulations are
level separation $\Delta$, Eq.(\ref{HS}),
dissipation rate $\gamma_0$ 
and cut-off frequency of the bath spectral function
(see Eqs.~(\ref{appB_defsigma})-(\ref{appB_defgam0}) for definitions),
inverse temperature $\beta$, Eq.(\ref{appB_BE}),
driving force number of modes $M$ and
bound on signal amplitude $\max\,\lvert c_k\rvert$, Eq.(\ref{epsilon}),
size and step of the FFT grid, and maximum number of
iterations employed by the optimization algorithm.
Control time $t_c$ is defined in each control task.
Below, all energies are presented in units of level separation $\Delta$,
time unit is $1/\Delta$.
Numerical values of the parameters are collected in Table~\ref{tab:par}.
We note that the parameters are favorable for the applicability
of the Redfield QME.

\begin{table}[htb]
    \centering
    \caption{Parameters for numerical simulation}
    \begin{tabular}{l@{\hskip 4em}r}
        \toprule
        Parameter & Value\\
        \hline
        $\Delta$ & $1$ \\ %0.2\\
        $\gamma_0$ & $0.5$ \\ %0.1\\
         $\omega_c$ & $5$ \\ %1.0 \\
        $\beta^{-1}$  & $1$ \\ %$1.0/\Delta$ \\
        $M$ & 4\\
        $\max|c_k|$ & $10$ \\ %2.0\\
        Size of the FFT grid & 80000 \\
        Energy step of the grid & $\pi/t_c$ \\
        Max number of iterations & 10\\
        Population size & 15\\
        \hline
    \end{tabular}
    \label{tab:par}
\end{table}

We now present results of simulations within the QME and NEGF 
for reset, heating, and cooling of qubit.

\subsection{Reset}
System starts in an arbitrary pure state and control goal is to bring it 
to a specific predefined pure state. As an example, we consider system evolution from
$\rho_i = (I + \sigma_z)/2$ to $\rho_f = (I-\sigma_x)/2$.
Here, $I$ is the identity matrix.

\begin{figure}[htbp]
    \centering
    \subfloat[\label{fig:reset10:a}]{
        \includegraphics[width=\columnwidth]{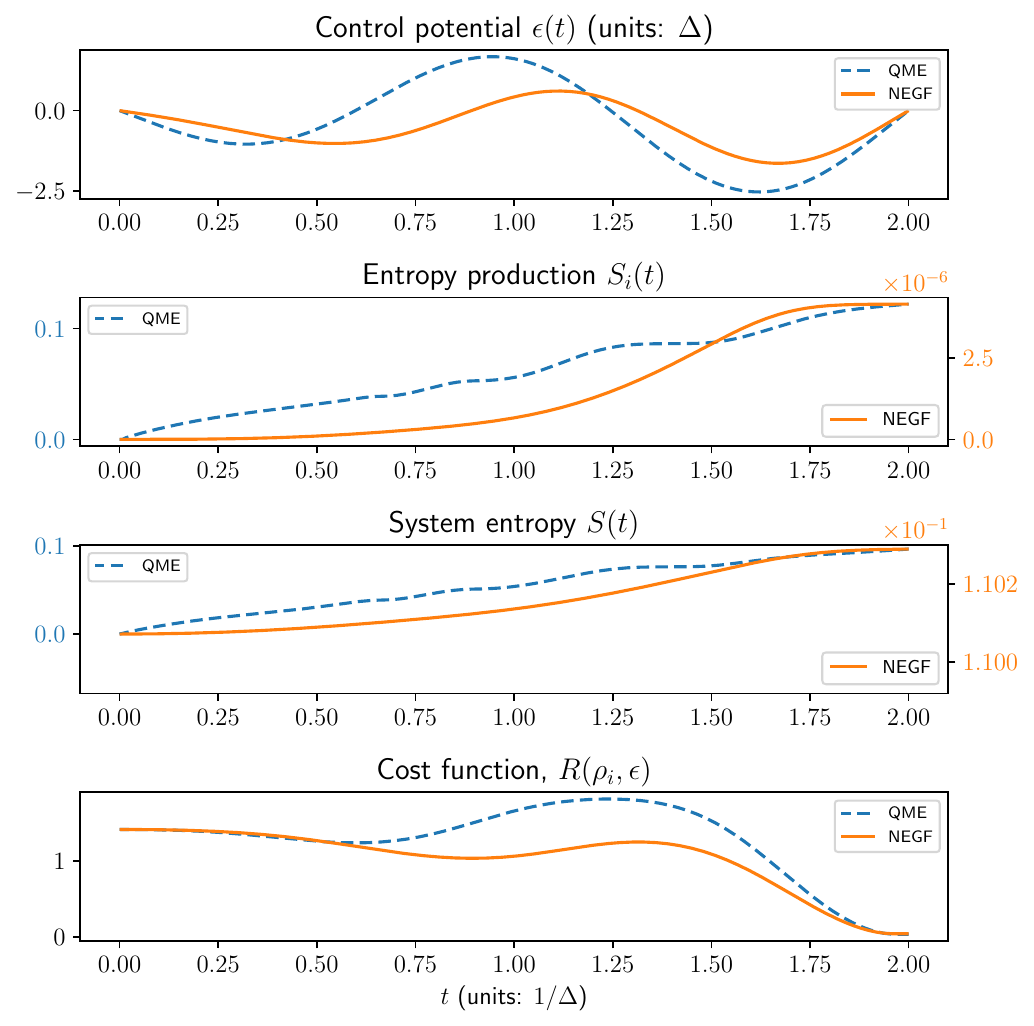}
    }\hfill
    \subfloat[\label{fig:reset10:b}]{
        \includegraphics[width=\columnwidth]{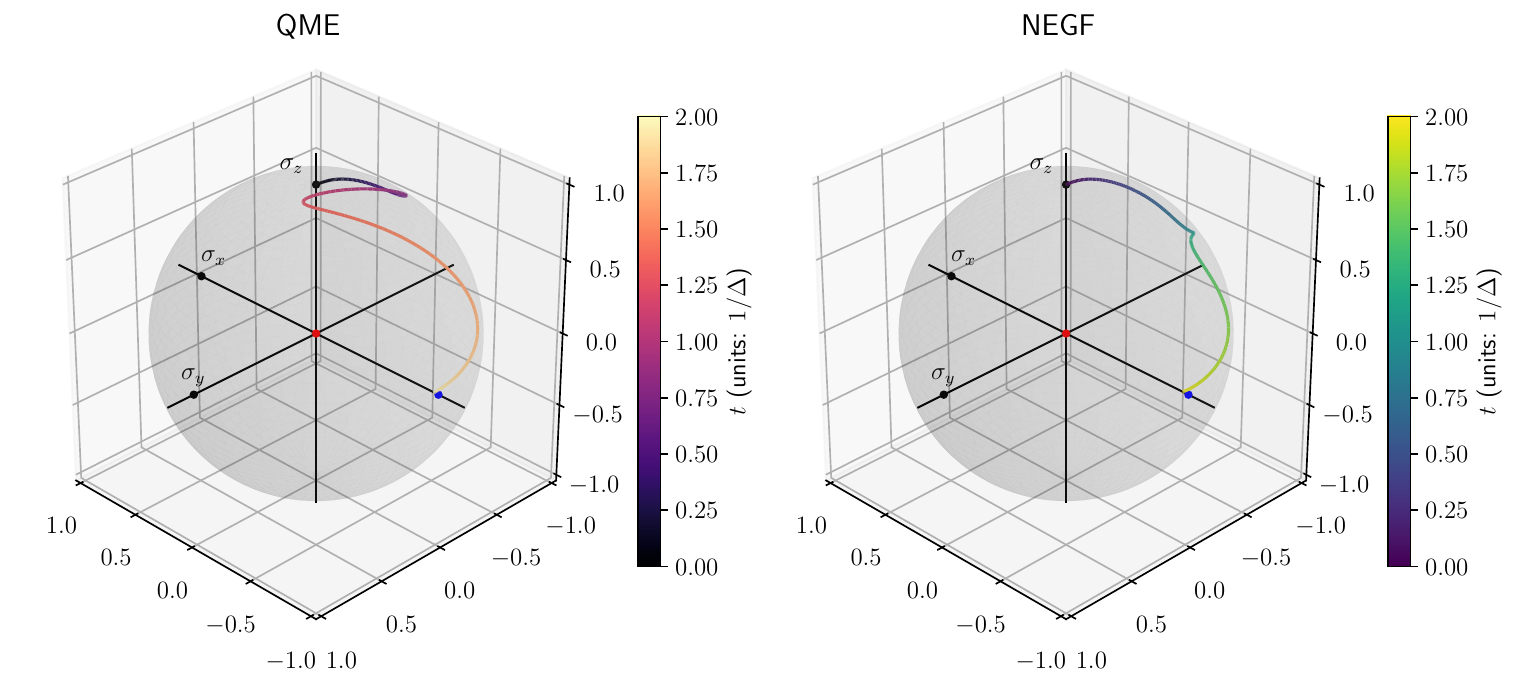}
    }\hfill
    \caption{(Color online) Reset of the qubit: forced evolution from an arbitrary initial state 
    ($\rho_i = (I + \sigma_z)/2$) to a predefined final state ($\rho_f = (I-\sigma_x)/2$).
    Panel (a) shows (top to bottom) optimal driving force potential, entropy production,
    entropy of the qubit, and cost function vs. time simulated within QME (dashed line, blue)
    and NEGF (solid line, orange). Panel (b) shows trajectories of the system
    within the two formulations.
    }
    \label{fig:reset10}
\end{figure}

Figure~\ref{fig:reset10} shows results of simulations performed within QME and 
NEGF for $t_c=2$.
The performance of optimization within the two formulations  is very similar.
Qualitative behavior of the qubit response is the same (see panel a)
with NEGF results demonstrating a slight time delay - a manifestation
of the non-Markov character of the evolution. As expected, entropy 
production is seen to level by the end of the driving. 
Both schemes are capable of reaching the control goal with the same accuracy.
Note that even trajectories (see panel b) are quite similar in the short time
for the two methods. This similarity is observed also in the other 
control tasks discussed below and is due to the initial condition 
where qubit and the bath are decoupled.
Such overall similarity of pure state to pure state transition
is not surprising: with coupling to driving force much stronger than 
dissipation (which is a prerequisite for ability to control the qubit),
dissipation almost doesn't play any role in a pure state evolution
(note that both trajectories are confined to outer surface of the sphere)
while unitary evolution is the same under both QME and NEGF
formulations.

%%%%%%%%%%%%%%%%%%%%%%%%%%%%%%%%%%%%%%%%

\subsection{Heating}
Heating task is the process starting from a pure (cold) state
evolving into the maximum entropy state, $\rho_f = I/2$.
We choose initial (cold) state to be $\rho_i = (I + \sigma_z)/2$ which corresponds
to $(0,0,1)$ vector on the sphere.

\begin{figure}[htbp]
    \centering
    \subfloat[\label{fig:heating30:a}]{
        \includegraphics[width=\columnwidth]{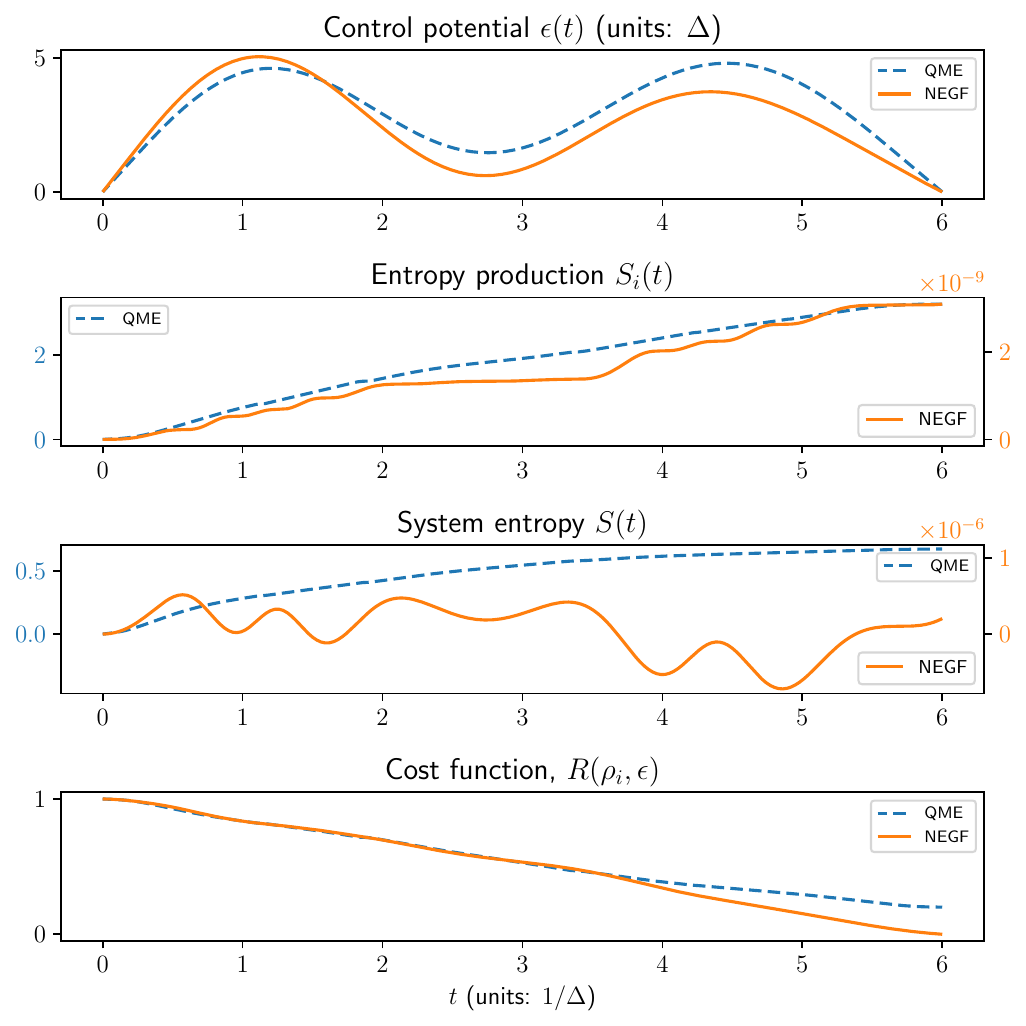}
    }\hfill
    \subfloat[\label{fig:heating30:b}]{
        \includegraphics[width=\columnwidth]{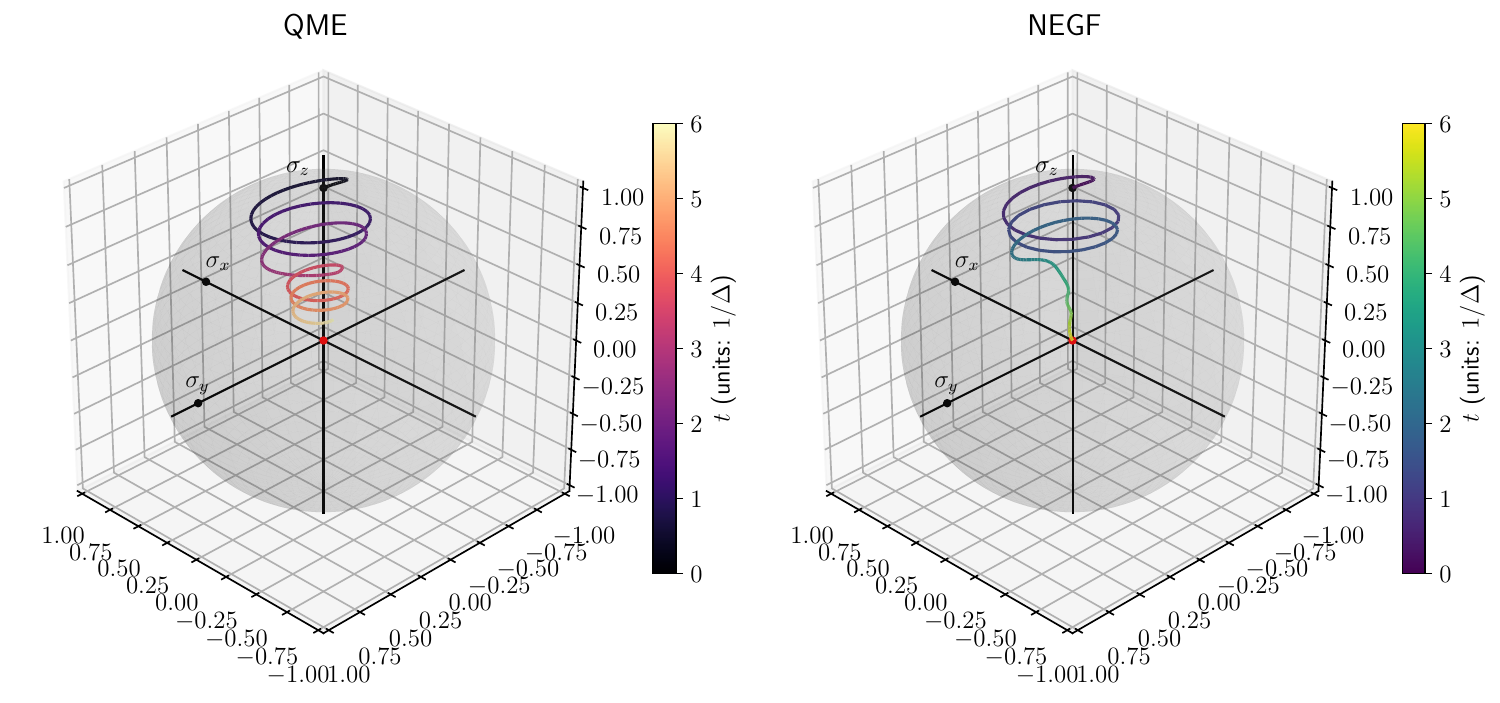}
    }\hfill
    \caption{(Color online) Heating of the qubit: forced evolution from an arbitrary initial state 
    ($\rho_i = (I + \sigma_z)/2$) to  the maximum entropy state ($\rho_f = I/2$).
    Panel (a) shows (top to bottom) optimal driving force potential, entropy production,
    entropy of the qubit, and cost function vs. time simulated within QME (dashed line, blue)
    and NEGF (solid line, orange). Panel (b) shows trajectories of the system
    within the two formulations.}
    \label{fig:heating30}
\end{figure}

Figure~\ref{fig:heating30} shows results of simulations performed within QME and 
NEGF for $t_c=6$.
Contrary to the previous task, evolution involving mixed states 
demonstrates differences between the results of  the two methods.
Indeed, change from pure to mixed state is impossible without dissipation.
The latter is treated very differently by QME (Markov) and NEGF (non-Markov)
formulations. This difference manifests itself in significant deviations
of trajectories in the region where the mixed character of the qubit is pronounced,
that is deep inside the sphere (see panel b). 

Note that NEGF simulation yields non-monotonic entropy change
with the possibility to have negative entropy values (see panel a). 
Both effects are due to dynamics of entanglement formation between
system and bath during evolution and are a manifestation of non-negligible
system-bath coupling (which is completely missed by the QME formulation).

Note also that for $t_c=6$ NEGF is faster in reaching the 
final state (see bottom graph in panel a). 
Longer driving ($t_c=12$, not shown) is necessary for the QME
to achieve the same goal. 

%%%%%%%%%%%%%%%%%%%%%%%%%%%%%%%%%%%%%%%%

\begin{figure}[t]
    \centering
    \subfloat[\label{fig:cooling100:a}]{
        \includegraphics[width=\columnwidth]{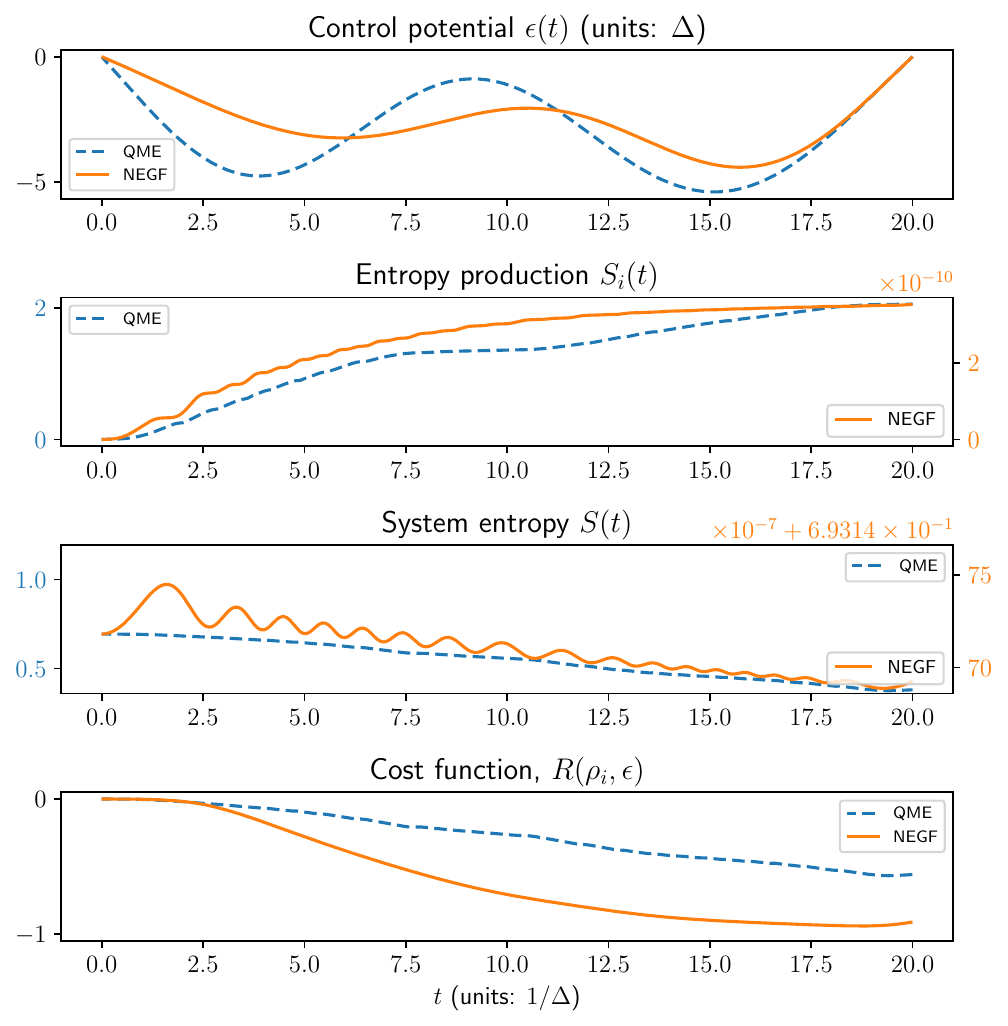}
    }\hfill
    \subfloat[\label{fig:cooling100:b}]{
        \includegraphics[width=\columnwidth]{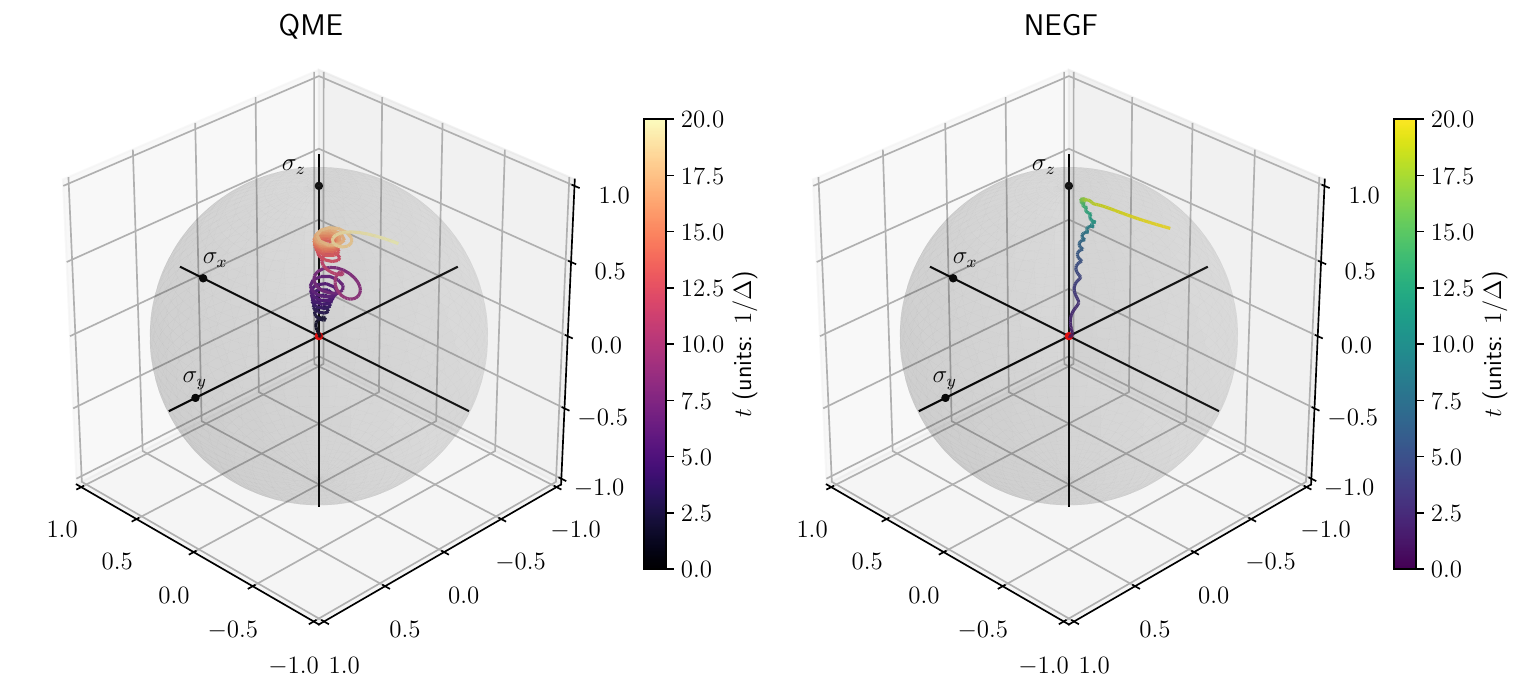}
    }\hfill
    \caption{(Color online) Cooling of the qubit: forced evolution from 
    the maximum entropy state ($\rho_i = I/2$) to a pure state ($\|\rho_f\| = 1$).
    Panel (a) shows (top to bottom) optimal driving force potential, entropy production,
    entropy of the qubit, and cost function vs. time simulated within QME (dashed line, blue)
    and NEGF (solid line, orange). Panel (b) shows trajectories of the system
    within the two formulations.}
    \label{fig:cooling100}
\end{figure}

\subsection{Cooling}
Cooling is a process opposite to heating:
initial state is the maximum entropy state
$\rho_i = I/2$ which evolves into a pure state $\rho_f$
where $\lvert\rho_f\rvert = 1$.

Figure~\ref{fig:cooling100} shows results of simulations performed within QME and 
NEGF for $t_c=100$. As with heating, the evolution involves a mixed state which
leads to pronounced differences between QME and NEGF results.
As previously,  NEGF results demonstrate non-monotonic behavior of 
system entropy (see panel a) and drastically different trajectory (panel b).
Again, NEGF is much more efficient
in reaching the final goal
(the optimized cost function value is $-0.64$ for QME and $-0.91$ for NEGF).
Note that simulations with longer driving ($t_c=40$, not shown)
also demonstrated the superiority of the NEGF formulation.
Moreover, increasing the number of iterations of the optimization cycle 
from $10$ to $100$ (not shown) didn't affect the inability of the QME 
to reach the final goal.
The difficulty of the qubit to escape from a low purity state to a high purity state
within QME (which is similar to the difficulty of the opposite process, as discussed 
above in heating task)  
is due to  approximate (adiabatic) treatment of the effect of the driving
on system dissipation.

%%%%%%%%%%%%%%%%%%%%%%%%%%%%%%%%%%%%%%%%%%%%%%%%%%%%%%%

\section{Conclusion} \label{conclude}
We examine the manipulation of a qubit coupled to a thermal bath using 
an external driving field. The optimization of the field (the time profile of the driving) 
aims to achieve a specific outcome (a particular state of the qubit) within a fixed 
duration of driving. We address three objectives: 
reset (evolution from an arbitrary pure state to a predefined pure state of the qubit), 
heating (evolution from a pure state to the maximum entropy state, $\rho_f = I/2$,
of the qubit), and cooling (evolution from the maximum entropy state, $\rho_f = I/2$,
to a pure state of the qubit).

Typically, studies on optimization in open quantum systems utilize 
the Redfield/Lindblad quantum master equation (QME) to characterize 
the system's dynamics. Among the various limitations of this method, 
the most significant for optimal control is the lack of consideration for 
the influence of time-dependent driving on the dissipator superoperator 
(in the standard formulation), or an approximate (adiabatic) approach 
to account for the effect of driving on dissipation (in the generalized formulation).

We utilize the NEGF, where the impact of driving on dissipation is modeled 
more accurately, and compare the predictions of the two methods. 
We find that in evolutions involving pure states (reset), the two formulations yield 
qualitatively similar results. This is because any optimization task entails a much 
stronger coupling to the driving field than to the bath. 
Consequently, the effect of dissipation on dynamics - where the two methods diverge -
is minor. 
Conversely, in evolutions involving mixed states (particularly for transitions between pure and mixed states, as seen in heating and cooling tasks), 
the two methods produce significantly different results, with the NEGF proving more effective in achieving the desired outcome. 
This is hardly surprising, as any transition between pure and mixed states 
necessitates the presence of dissipation.

We note that our simulations were conducted in a parameter regime conducive 
to employing the Redfield QME. However, the adiabatic approximation used 
to address the impact of time-dependent driving on dissipation within 
the method leads to challenges in accurately describing mixed state evolution. 
This highlights the advantages of non-Markov Green's function techniques and, hopefully, will establish them as a preferred tool for studies on optimization 
in open quantum systems.

%%%%%%%%%%%%%%%%%%%%%%%%%%%%%%%%%%%%%%%%%%%%%%%%%%%%%%%

\begin{acknowledgments}
We thank Ronnie Kosloff and Shimshon Kallush for helpful discussions.
This material is based upon work supported by the National Science Foundation under Grant No. CHE-2154323.
\end{acknowledgments}

%%%%%%%%%%%%%%%%%%%%%%%%%%%%%%%%%%%%%%%%%%%%%%%%%%%%%%%
%%%%%%%%%%%%%%%%%%%%%%%%%%%%%%%%%%%%%%%%%%%%%%%%%%%%%%%

\appendix
\section{Derivation of Eq.(\ref{QME})} \label{appA}
Derivation of the time-dependent QME (\ref{QME}) follows standard 
procedure for the Redfield QME~\cite{breuer_theory_2003}
with the modification accounting for time-dependence of the driving field
on dissipator as is described in  Ref.~\onlinecite{kallush_controlling_2022}.

Specifically, we start from the Redfield equation 
\begin{align}
 \label{appA_qme0}
  &\frac{d}{d t}\hat{\tilde\rho}_S(t) = 
  \\ &
  -\int_0^\infty ds\,
  \mbox{Tr}_B\bigg\{\left[\hat{\tilde H}^{SB}(t), \left[\hat{\tilde H}^{SB}(t-s), \hat{\tilde\rho}_S(t)\otimes\hat\rho_B^{eq}\right]\right]\bigg\}
\nonumber
\end{align}
Here, tilde indicates interaction picture, 
$\hat{\tilde\rho}_S(t)$ is the reduced density matrix of the qubit, and
$\hat\rho_B^{eq}$ is the equilibrium density operator of the thermal bath. 
$\mbox{Tr}_B\left\{\ldots\right\}$ is trace over the bath degrees of freedom.

System-bath coupling (\ref{HSB}) in interaction picture is
\begin{equation}
\label{appA_HSB_int}
\begin{split}
\hat{\tilde H}^{SB}(t) &\equiv \hat U_S^\dagger(t,0)\,
\frac{i}{2}\left(\hat d_2^\dagger\hat d_1-\hat d_1^\dagger\hat d_2\right)\,
\hat U_S(t,0)
\\ & \otimes \hat U_B^\dagger(t,0)\,\hat B\,\hat U_B(t,0)
\end{split}
\end{equation}
where $\hat U_S(t,0)$ is the system free evolution operator which satisfies
\begin{equation}
\label{US_EOM}
\frac{\partial}{\partial t}\hat U_S(t,0) = -i\hat H^S(t)\,\hat U_S(t,0),
\end{equation} 
$\hat B\equiv\sum_\alpha g_\alpha\left(\hat a_\alpha+\hat a_\alpha^\dagger\right)$,
and $\hat U_B(t,0)$ describes evolution of free bath which is assumed
to be in thermal equilibrium.

Solving numerically free evolution (\ref{US_EOM})
one can introduce instantaneous eigenproblem for the operator 
$\hat U_S(t,0)$
\begin{equation}
\hat U_S(t,0)\lvert S(t)\rangle = e^{-i u_S(t)}\lvert S(t)\rangle
\end{equation} 
Eigenvectors are used to define instantaneous jump operators 
\begin{equation}
\label{appA_defF}
\hat F_{S_1S_2}(t)\equiv \lvert S_1(t)\rangle\langle S_2(t)\rvert
\end{equation}
which evolve as
\begin{align}
\hat U_{S}^\dagger(t,0)\hat F_{S_1S_2}(t)\hat U_S(t,0) &=
e^{i\left[u_{S_1}(t)-u_{S_2}(t)\right]} \hat F_{S_1S_2}(t)
\nonumber \\ &\equiv e^{-i\theta_{S_1S_2}(t)}\hat F_{S_1S_2}(t)
\end{align}
These jump operators can be used to represent free evolution of any 
operator acting on system degrees of freedom only. 
In particular,
\begin{equation}
\label{appA_sy_evolv}
\begin{split}
&\hat U_S^\dagger(t,0)\,
\frac{i}{2}\left(\hat d_2^\dagger\hat d_1-\hat d_1^\dagger\hat d_2\right)\,
\hat U_S(t,0) 
\\ & \qquad\qquad\qquad =
\sum_m \xi_m(t)e^{-i\theta_m(t)}\hat F_{m}(t)
\end{split}
\end{equation}
where $m\equiv (S_1,S_2)$ indicates transition between states of the system
and $\xi_m(t)$ are coefficients of expansion.

Using (\ref{appA_sy_evolv}) in (\ref{appA_qme0}) leads to
\begin{align}
\label{appA_qme1}
&\frac{d}{dt}\hat{\tilde\rho}_S(t) =
\sum_{m,m'}\int_0^\infty ds\, \xi_m^{*}(t)\,\xi_{m'}(t-s)\,
e^{i\left[\theta_m(t)-\theta_{m'}(t-s)\right]}
\nonumber \\ &\times
\left[
\hat F_{m'}(t-s)\hat{\tilde\rho}_S(t)\hat F_m^\dagger(t) - 
\hat F_m^\dagger(t)\hat F_{m'}(t-s)\hat{\tilde\rho}_S(t)
\right]
\nonumber \\ &\times
\left\langle \hat{\tilde B}(t)\,\hat{\tilde B}(t-s)\right\rangle_{eq}
+\mbox{H.c.}
\end{align}
where we used Hermitian property of $\hat s_y$, $\hat B$, Hamiltonian,
 and density operator,
and where
\begin{equation}
\left\langle \hat{\tilde B}(t)\,\hat{\tilde B}(t-s)\right\rangle_{eq} \equiv
\mbox{Tr}_B\left\{\hat{\tilde B}(t)\,\hat{\tilde B}(t-s)\,\hat\rho_B^{eq}\right\}
\end{equation}

Standard derivation of the Redfield QME implies infinitely fast bath,
this allows to approximate
\begin{equation}
\xi_{m'}(t-s)\approx \xi_{m'}(t)\quad\mbox{and}\quad
\hat F_{m'}(t-s)\approx\hat F_{m'}(t)
\end{equation}
Expanding phase factor in (\ref{appA_qme1}) to linear term in time
and employing  rotating wave approximation,
\begin{equation}
 \theta_m(t)-\theta_{m'}(t-s)\approx \delta_{m,m'}\omega_m s,
\end{equation}
leads to
\begin{align}
\label{appA_qme2}
&\frac{d}{dt}\hat{\tilde\rho}_S(t) = 
\\ &
\sum_m\Gamma_m(t)
\left[\hat F_m(t)\hat{\tilde\rho}_S(t)\hat F_m^\dagger(t)-
\hat F_m^\dagger(t)\hat F_{m}(t)\hat{\tilde\rho}_S(t)\right]
+\mbox{H.c.}
\nonumber
\end{align}
where
\begin{equation}
\label{appA_defGamma}
\Gamma_m(t)\equiv \lvert\xi_m(t)\rvert^2\int_0^\infty ds\,
e^{i\omega_m s}\left\langle\hat{\tilde B}(s)\,\hat{\tilde B}(0)\right\rangle_{eq}
\end{equation}
Expression (\ref{appA_qme2}) is Eq.(\ref{QME}).

%%%%%%%%%%%%%%%%%%%%%%%%%%%%%%%%%%%
%%%%%%%%%%%%%%%%%%%%%%%%%%%%%%%%%%%

%%%%%%%%%%%%%%%%%%%%%%%%%%%%%%%%%%%%%%%%%%%%%%%%%%%%%%%

\section{Qubit self-energy due to coupling to thermal bath} \label{appB}
It is convenient to express system-bath coupling, Eq.(\ref{HSB})
in a more general form
\begin{align}
    \hat H^{SB} &= \sum_{i,j} v_{ij} d_i^\dagger d_j
        \sum_\alpha g_\alpha(\hat a_\alpha + \hat a_\alpha^\dagger) 
\end{align}
where $v_{ij} = \frac{i}{2}(\delta_{i,2}\delta_{j,1} - \delta_{i,1}\delta_{j,2})$.
The coupling is considered to be a perturbation. Below, it is taken into account
employing diagrammatic expansion.

Rewriting expression for the single-particle Green's function, Eq.(\ref{GF}), 
in the interaction picture,
\begin{equation}
 \label{appB_GF}
    G_{ij}(\tau_1, \tau_2) = 
     -i\left\langle T_c\, \hat{\tilde d}_{i}(\tau_1)\,\hat{\tilde d}_{j}^\dagger(\tau_2)\, 
    e^{-i\int_c d\tau\,\hat{\tilde H}^{SB}(\tau)}\right\rangle,
\end{equation}
expanding evolution operator up to second order in $\hat{\tilde H}^{SB}$,
applying the Wick's theorem, and dressing the diagrams yields 
the Dyson equation - integral form of the Kadanoff-Baym equation 
(\ref{KB})
\begin{align}
    & G_{ij}(\tau_1, \tau_2) = G_{ij}^{(0)}(\tau_1, \tau_2)
    \\ &
    + \sum_{l,m}\int_c d\tau\int_c d\tau'\, G_{il}^{(0)}(\tau_1, \tau)\,
    \Sigma_{lm}(\tau, \tau')\, G_{mj}(\tau', \tau_2)
    \nonumber
\end{align}
where $G^{(0)}$ is the single-particle Green's function in the absence of
of the system-bath coupling and $\Sigma$ is self-energy due to the coupling. 
Explicit expression of the self-energy within the second order epxansion
(the Hartree-Fock approximation) is  
\begin{align}
\label{appB_SE}
    \Sigma_{lm}(\tau, \tau') &= 
    \underbrace{\Sigma_{lm}^\mathrm{H}(\tau, \tau')}_\mathrm{Hartree} + 
    \underbrace{\Sigma_{lm}^\mathrm{F}(\tau, \tau')}_\mathrm{Fock}
    \\
    \label{appB_SEH}
    \Sigma_{lm}^\mathrm{H}(\tau, \tau') 
        &= -i \delta(\tau, \tau')\sum_{m'l'}\int_c d\tau'' v_{lm}v_{l'm'}
        \\ &\times
        G_{m'l'}(\tau'',\tau''_+)[\sigma(\tau'', \tau) + \sigma(\tau, \tau'')]
        \nonumber \\
        \label{appB_SEF}
    \Sigma_{lm}^\mathrm{F}(\tau, \tau') &= i\sum_{m'l'}v_{lm'}v_{l'm}
    \\ & \times
        G_{m'l'}(\tau, \tau')
        [\sigma(\tau',\tau) + \sigma(\tau,\tau')]
     \nonumber
\end{align}
where 
\begin{align}
\label{appB_defsigma}
    \sigma(\tau,\tau') 
        &= -i\sum_{\alpha}g_{\alpha}^2
        \left\langle T_c\, \hat a_{\alpha}(\tau)\,\hat  a_{\alpha}^\dagger(\tau')\right\rangle_0
\end{align}
Here, $\langle\ldots\rangle_0$ indicates free evolution of the bath.

\iffalse
\begin{figure}[htb]
    {\centering\includegraphics[width=0.8\linewidth]{fig5.png}}
    \caption{Feynman diagrams of the self-energy (\ref{appB_SE}). 
    Shown are (a) Hartree $\Sigma^\mathrm{H}$, Eq.(\ref{appB_SEH}), and 
    (b) Fock $\Sigma^\mathrm{F}$, Eq.(\ref{appB_SEF}), contributions.}
\end{figure}
\fi

\begin{figure}[htbp]
{\centering
\hspace*{1.5cm}
\subfloat[][]{
\begin{tikzpicture}
\begin{feynman}
\vertex (x1);
\vertex [above=1cm of x1] (x2);
\vertex [above=1cm of x2] (x3);
\diagram*[layered layout] {
(x1) -- [photon] (x2),
(x2) -- [fermion, half left] (x3),
(x3) -- [fermion, half left] (x2),
};
\end{feynman}
\end{tikzpicture}
}
\hfill
\subfloat[][]{
\begin{tikzpicture}
\begin{feynman}
\vertex (x1);
\vertex [right=2cm of x1] (x2);
\diagram*[layered layout] {
(x1) -- [fermion] (x2),
(x1) -- [photon, half left] (x2),
};
\end{feynman}
\end{tikzpicture}
}
\hspace*{1.5cm}
}
\caption{Feynman diagrams of the self-energy (\ref{appB_SE}).
Shown are (a) Hartree $\Sigma^\mathrm{H}$, Eq.(\ref{appB_SEH}), and
(b) Fock $\Sigma^\mathrm{F}$, Eq.(\ref{appB_SEF}), contributions.}
\end{figure}
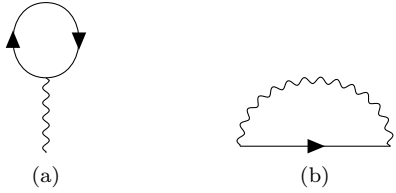

In simulations, we'll need lesser and greater projections of (\ref{appB_defsigma}).
To obtain those we start from energy domain where
\begin{equation}
\label{appB_sigmaltgt}
\begin{split}
    \sigma^<(\omega) &= -i\gamma(\omega)N(\omega)
    \\
    \sigma^>(\omega) &= -i\gamma(\omega)[N(\omega) + 1]
    \end{split}
\end{equation}
and employ FFT. 
In (\ref{appB_sigmaltgt})
\begin{align}
    \label{appB_defgam0}
    \gamma(\omega) &= \gamma_0\left( \frac{\omega}{\omega_c} \right)^2
        e^{2\left( 1 - \frac{\omega}{\omega_c} \right)}\\
    \label{appB_BE}
    N(\omega) &= \frac{1}{e^{\beta\omega} - 1}
\end{align}
are the dissipation rate and Bose-Einstein thermal distribution.

We note that because thermal bath (\ref{appB_sigmaltgt})-(\ref{appB_BE})
does not support zero frequency modes, only Fock self-energy,
Eq.(\ref{appB_SEF}), will contribute. 

%%%%%%%%%%%%%%%%%%%%%%%%%%%%%%%%%%%%%%%%%%%%%%%%%%%%%%%

\section{NEGF expressions for the heat flux and entropy production rate}\label{appC}
Here we derive expressions for heat flux and entropy production rate
of a qubit coupled to thermal bath.

We start from considering the particle (phonon) flux on the system-bath interface.
The flux is defined as minus rate of change of population in the bath
\begin{equation}
I_B(t) = -\frac{d}{d t}\sum_\alpha\langle{\hat a_\alpha^\dagger(t) \hat a_\alpha(t)}\rangle
\end{equation}
For the Hamiltonian (\ref{H})-(\ref{HSB}) it can be expressed in terms of mixed Green's function 
\begin{equation}
\label{appC_IB_1}
    I_B(t) = 2\,\mbox{Re} \left[\sum_\alpha g_\alpha\, G_{\alpha S}^<(t, t) \right]
\end{equation}
where the Green's function is lesser projection of  
\begin{align}
\label{appC_defGaS}
    G_{\alpha S}(\tau_1, \tau_2) &\equiv -i\langle T_c\, \hat a_\alpha(\tau_1)\, \hat S^\dagger(\tau_2)\rangle
\end{align}
Here,
\begin{equation}
\hat S\equiv \frac{i}{2}\left(\hat d_2^\dagger\hat d_1-\hat d_1^\dagger\hat d_2\right)
\end{equation}
Using the Dyson equation, mixed Green’s function is expressed in terms of 
free bath and system evolutions as
\begin{equation}
\label{appC_GaS}
    G_{\alpha S}(\tau_1, \tau_2) = \int_c d\tau F^{(0)}_\alpha(\tau_1, \tau)\, g_\alpha\, 
    G_{SS}(\tau, \tau_2)
\end{equation}
where 
\begin{align}
    F^{(0)}_\alpha(\tau_1, \tau_2) &= -i\langle T_c\, \hat a_\alpha(\tau_1)\,
    \hat a_\alpha^\dagger(\tau_2)\rangle_0
    \\ 
    G_{SS}(\tau_1, \tau_2) &= -i\langle T_c\, \hat S(\tau_1)\, \hat S^\dagger(\tau_2)\rangle
\end{align}
are Green's functions describing evolution of free bath phonon and full system excitation, respectively.

Substituting lesser projection of (\ref{appC_GaS}) into (\ref{appC_IB_1})
leads to
\begin{align}
\label{appC_IB_2}
    I_B(t) &= -2\,\mbox{Re}\int_{0}^t d t'\, \int\frac{d\omega}{2\pi}\,
    e^{-i\omega(t-t')}
    \\ &\qquad\times
    \left[\sigma^<(\omega)\, G_{SS}^>(t', t) - 
       \sigma^>(\omega)\, G_{SS}^<(t', t)
    \right]
    \nonumber \\ 
    \label{appC_iB}
    &\equiv \int \frac{d\omega}{2\pi}\, i_B(t, \omega)
\end{align}
Here, $i_B(t,\omega)$ is the energy resolved particle (phonon) flux
and $\sigma^{\lessgtr}$ is Fourier transform of the lesser/greater 
projection of
\begin{equation}
\sigma(\tau_1,\tau_2)\equiv\sum_\alpha g_\alpha^2\, F^{(0)}_\alpha(\tau_1,\tau_2)
\end{equation}
Explicit expressions of the projections are given in Eq.(\ref{appB_sigmaltgt}).

In terms of energy resolved particle flux (\ref{appC_iB}), 
heat flux (which for the thermal bath is equivalent to energy flux) is 
\begin{equation}
\label{appC_QB}
{\dot Q}_B(t) = \int_0^\infty\frac{d\omega}{2\pi}\, \omega\, i_B(t,\omega)
\end{equation}

To derive expression for entropy production, we express particle flux (\ref{appC_IB_2})
as difference between incoming thermal and outgoing non-thermal fluxes.
In writing this expression we take into account that in 1D velocity
exactly cancels with density of states~\cite{buttiker_scattering_1992}. Thus, 
\begin{equation}
\label{appC_IB_3}
I_B(t) = \int\frac{d\omega}{2\pi}\,\left[\phi^{in}(\omega)-\phi^{out}(t,\omega)\right]
\end{equation}
Because expression for thermal population is known, 
\begin{equation}
\phi^{in}(\omega) = N(\omega),
\end{equation}
Eqs.~(\ref{appC_iB}) and (\ref{appC_IB_3}) yield expression for 
outgoing non-thermal flux
\begin{equation}
\phi^{out}(t,\omega) = \phi^{in}(\omega)-i_B(t,\omega)
\end{equation}
Rate of entropy change is introduced as a difference between
incoming and outgoing entropy fluxes
\begin{equation}
\label{appC_dS}
\frac{dS}{dt}=\int\frac{d\omega}{2\pi}\left(\sigma[\phi^{in}(\omega)]-\sigma[\phi^{out}(t,\omega)]\right)
\end{equation}
where
\begin{equation}
 \sigma[\phi]\equiv -\phi\,\ln\phi - (1-\phi)\,\ln(1-\phi)
\end{equation}
is the von Neumann expression for entropy in the bath.
Finally, using (\ref{appC_QB}) and (\ref{appC_dS}) in (\ref{2law})
yields expression for entropy production~\cite{zhou_quantum_2024}
\begin{align}
\label{appC_dSi}
    &{\dot S}_i(t) = \int_0^\infty\frac{d\omega}{2\pi}\bigg\{
        \phi^{out}(t, \omega)[\ln\phi^{out}(t, \omega) - \ln\phi^{in}(\omega)]
        \\ &
        - \big(1 + \phi^{out}(t, \omega)\big)\big[\ln\big(1+\phi^{out}(t, \omega)\big) - 
            \ln\big(1 + \phi^{in}(\omega)\big)\big]
    \bigg\}
    \nonumber
\end{align}
Note that consideration above is done in the diagonal approximation.
General formulation can be found in Ref.~\onlinecite{zhou_quantum_2024}.

%%%%%%%%%%%%%%%%%%%%%%%%%%%%%%%%%%%%%%%%%%%
%%%%%%%%%%%%%%%%%%%%%%%%%%%%%%%%%%%%%%%%%%%

%\bibliography{j_abbrev,control_qm}


\end{document}